\documentclass[aps,prd,groupedaddress,twocolumn,floatfix,nofootinbib,showpacs,showkeys]{revtex4-1}

\usepackage{graphicx}
\usepackage{dcolumn}
\usepackage{bm}
\usepackage{amsmath}
\usepackage{epstopdf}
\usepackage[flushleft]{threeparttable}

\def\mnras{MNRAS}
\def\apj{Astroph. J.}
\def\apjl{Astroph. J.}

\def\aap{Astron. Astroph.}

\begin{document}

\title{On the last stable orbit around rapidly rotating neutron stars}
\author{F.~Cipolletta,$^{1}$ C.~Cherubini,$^{2,3}$ S.~Filippi,$^{2,3}$ J.~A.~Rueda,$^{1,4,5}$, R.~Ruffini$^{1,4,5}$}
\email{cipo87@gmail.com; c.cherubini@unicampus.it;\\s.filippi@unicampus.it; jorge.rueda@icra.it; ruffini@icra.it}
\affiliation{$^1$Dipartimento di Fisica and ICRA, Sapienza Universit\`a di Roma, P.le Aldo Moro 5, I--00185 Rome, Italy}
\affiliation{$^2$Unit of Nonlinear Physics and Mathematical Modeling, University Campus Bio-Medico of Rome, Via A.~del Portillo 21, I--00128 Rome, Italy}
\affiliation{$^3$International Center for Relativistic Astrophysics-ICRA, University Campus Bio-Medico of Rome, Via A.~del Portillo 21, I--00128 Rome, Italy}
\affiliation{$^4$ICRANet, Piazza della Repubblica 10, I--65122 Pescara, Italy}
\affiliation{$^5$ICRANet-Rio, Centro Brasileiro de Pesquisas F\'isicas, Rua Dr. Xavier Sigaud 150, Rio de Janeiro, RJ, 22290--180, Brazil }

\date{\today}

\begin{abstract}
We compute the binding energy and angular momentum of a test-particle at the last stable circular orbit (LSO) on the equatorial plane around a general relativistic, rotating neutron star (NS). We present simple, analytic, but accurate formulas for these quantities that fit the numerical results and which can be used in several astrophysical applications. We demonstrate the accuracy of these formulas for three different equations of state (EOS) based on nuclear relativistic mean-field theory models and argue that they should remain still valid for any NS EOS that satisfy current astrophysical constraints. We compare and contrast our numerical results with the corresponding ones for the Kerr metric characterized by the same mass and angular momentum.
\end{abstract}



\maketitle

\section{Introduction}\label{sec:1}

It is well-known that the knowledge of the properties of the circular orbits of test-particles, e.g. energy and angular momentum around compact objects such as neutron stars (NSs) and black holes (BHs) are of paramount importance for the understanding of several astrophysical scenarios such as the accretion processes in binary X-ray sources \cite{1975ASSL...48.....G}. 

The precise knowledge of NS properties is essential for the correct description of the NS structure evolution during the accretion process. This is particulary relevant in the evolution of accreting NSs in X-ray binaries leading to the NS spin-up and final formation of the millisecond \emph{recycled} pulsars \cite{1982Natur.300..728A}. It is by now clear that the inclusion in the accretion process of subtle effects such as the NS binding energy \cite{2011MNRAS.413L..47B}, and the precise energy and angular momentum transferred to the NS including general relativistic effects and the NS interior compression \cite{2011A&A...536A..87B,2011A&A...536A..92B}, can have an impact in the determination of the correct evolutionary scenario and therefore in the determination of the binary progenitors of millisecond pulsars (see, e.g., Refs.~\cite{2011MNRAS.416.2130T,2012MNRAS.425.1601T,2016A&A...586A.109F}).

On the other hand, it has been shown that such an information becomes also relevant within the induced gravitational paradigm of gamma-ray bursts (GRBs), where a hypercritical accretion process is triggered onto a NS by the supernova explosion of a binary companion carbon-oxygen core \cite{2012ApJ...758L...7R, 2014ApJ...793L..36F, 2015ApJ...812..100B, 2015PhRvL.115w1102F,2016ApJ...833..107B}. In contrast to binary X-ray sources in which the NS accretes matter from a companion at sub-Eddington rates $\dot{M}\equiv dM/dt \lesssim 10^{-8}~M_\odot$~y$^{-1}$, hence evolving quietly on very long time-scales $t_{\rm acc}\equiv M/\dot{M} \gtrsim 10^8$~y, the aforementioned hypercritical accretion process in GRBs leads to a NS which evolves in time-scales as short as $t_{\rm acc}= M/\dot{M} \sim 10^2$~s. In such a short time-interval, the NS can reach either the mass-shedding or the secular axisymmetric instability with consequent gravitational collapse to a BH (see, e.g., Refs.~\cite{2014ApJ...793L..36F, 2015ApJ...812..100B,2016ApJ...833..107B}). 

It is clear that the description of processes similar to the above one needs the knowledge of the properties of the NS interior, of its exterior spacetime, and of the circular orbits around it. The aim of this article is to provide these ingredients. 

Uniformly rotating NS equilibrium configurations form a two-parameter family of solutions characterized by baryonic mass $M_b$ and angular momentum $J$. We can write the evolution of a uniformly rotating NS gravitational mass $M$ as:
\begin{equation}
	\dot{M}=\left(\frac{\partial M}{\partial M_b}\right)_J\dot{M}_b+\left(\frac{\partial M}{\partial J}\right)_{M_b}\dot{J},
	\label{eq:Mns_evol}
\end{equation} 
where $\dot{M}_b$ and $\dot{J}$ are the amount of baryonic mass and angular momentum being transferred to the NS per-unit-time, namely the mass accretion rate and torque acting onto the NS. The two above partial derivatives have to be obtained from the relation $M(M_b,J)$ which is obtained numerically. We have recently found in Ref.~\cite{2015PhRvD..92b3007C} that, independent on the nuclear equation of state (EOS), such a relation for uniformly rotating NSs is well fitted by:
\begin{equation}
	\frac{M_b}{M_\odot}=\frac{M}{M_\odot}+\frac{13}{200}\left( \frac{M}{M_\odot} \right)^2\left( 1-\frac{1}{130}j^{1.7} \right),
	\label{eq:MbMnsjns}
\end{equation}
where $j \equiv c J /(G M_{\odot}^2)$. This relation has been shown to be very accurate also in the description of the binding energy of other nuclear EOS models including hyperonic and hybrid ones \cite{2017NewA...54...61B}.

The total energy released in an accretion process is given by the amount of gravitational energy gained by the material in its way to the NS surface and that is not spent in increasing the gravitational binding energy of the NS, namely (see, e.g., \cite{2000AstL...26..772S,2015ApJ...812..100B}):
\begin{eqnarray}
&&L_{\rm acc}= (\dot{M}_b - \dot{M})c^2 \nonumber \\
&&=\dot{M}_b c^2 \left[1-\left(\frac{\partial M}{\partial J}\right)_{M_b}\,\frac{\dot{J}}{\dot{M}_b} -\left(\frac{\partial M}{\partial M_b}\right)_{J}\right],
\label{eq:Lacc}
\end{eqnarray}
where we have used Eq.~(\ref{eq:Mns_evol}).

If the accretion of matter comes from a disk-like structure, such a total radiated energy $L_{\rm acc}$ is given by the sum of the energy radiated in the disk, $L_{\rm disk}$, and the energy radiated at the NS surface when the material is incorporated to the star, $L_s$, i.e.
\begin{equation}\label{eq:Lacc2}
L_{\rm acc} = L_s + L_{\rm disk}.
\end{equation}

In the case when the magnetic field effects can be neglected, the inner boundary of an accretion disk around a compact object is assumed to be given by the radius of the last stable circular orbit (hereafter LSO) of a test particle of mass $\mu\ll M$. Thus, the knowledge of the energy and angular momentum of a test particle at the LSO is essential for the determination of the evolution of the NS during the accretion process.  We denote hereafter by $\tilde{E}\equiv E/\mu$ and $\tilde{L}\equiv L/(G \mu M/c)$ the energy per-unit-mass and dimensionless angular momentum of a particle at the LSO.

From energy and angular momentum conservation we have that the mass-energy and angular momentum transferred to the NS from a particle infalling from the LSO are (see, e.g., Ref.~\cite{2000AstL...26..772S}):
\begin{eqnarray}
\dot{M}c^2 &=& \tilde{E} \dot{M}_b c^2 - L_s\label{eq:Mdot}\\
\dot{J} &=& \dot{M}_b \tilde{L} \frac{G M}{c}.\label{eq:Jdot}
\end{eqnarray}
Eqs.~(\ref{eq:Mns_evol})--(\ref{eq:Jdot}) lead therefore to the surface luminosity:
\begin{equation}\label{eq:Ls}
L_s=\dot{M}_b c^2 \left[\tilde{E}-\left(\frac{\partial M}{\partial J}\right)_{M_b} \tilde{L}\frac{G M}{c} -\left(\frac{\partial M}{\partial M_b}\right)_{J}\right],
\end{equation}
and to the disk luminosity:
\begin{equation}\label{eq:Ldisk}
L_{\rm disk} \equiv L_{\rm acc} - L_s= \dot{M}_b c^2 \left(1- \tilde{E}\right).
\end{equation}

From Eqs.~(\ref{eq:Mns_evol}--\ref{eq:Jdot}) one can compute the time evolution of the mass and angular momentum of the NS in an accretion process, providing we know how $\tilde{E}$ and $\tilde{L}$ depend on the gravitational (or on the baryonic mass) and angular momentum of the NS. At the same time, Eqs.~(\ref{eq:Ls}) and (\ref{eq:Ldisk}) give us, respectively, the surface and disk luminosities which are important from the observational point of view. It is worth to mention that the contribution of $L_s$ and $L_{\rm disk}$ to the total radiated energy can be comparable depending on the angular momentum \cite{2000AstL...26..772S}.

In this article we present simple but accurate fitting formulas of $\tilde{E}(M,J)$ and $\tilde{L}(M,J)$ both for co-rotating and counter-rotating orbits around rotating NS and are valid for any rotation rate within the NS region of stability bounded by mass-shedding and secular axisymmetric instability limits.

We show below that the aforementioned formulas for $\tilde{E}(M,J)$ and $\tilde{L}(M,J)$ are shown to be the same for three different nuclear EOS based on relativistic mean-field theory, suggesting a possible \emph{universal} character. We elaborate on this concept and show that current astrophysical constraints imply that, indeed, our formulas should remain valid for other astrophysically relevant set of EOS and for the relevant NS masses leading to an LSO located outside the NS surface.

Despite the complexity of NSs and the still debated EOS governing their interior physics, there have been discovered features which seem to be EOS-independent such as the relation between the moment of inertia, Love number and quadrupole moment, i.e. the $I$-Love-$Q$ relation \cite{2013Sci...341..365Y,2013PhRvD..88b3009Y}, and the NS binding energy shown in Eq.~(\ref{eq:MbMnsjns}) \cite{2015PhRvD..92b3007C}. We show in this work that indeed also the energy and angular momentum of the LSO around rotating NSs are very weakly EOS-dependent properties in the limits established by current astrophysical constraints. All the above allow the construction of a set of analytic and/or semi-analytic set of NS properties that can be used in a variety of NS astrophysical scenarios as the accretion process exemplified above.

The article is organized as follows. In section~\ref{sec:2} we compute the interior and exterior spacetime geometry of uniformly rotating NSs. The general formulation of the problem of circular orbits is recalled in section~\ref{sec:3}. Then, in section~\ref{sec:4} we compute the configurations for which there exists a LSO outside the NS surface. In section~\ref{sec:5} we focus on those configurations and compute the binding energy and angular momentum of the LSO. Finally, we shall present  simple but very accurate fitting formulas for these quantities. 

\section{Neutron star structure and spacetime geometry}\label{sec:2}

We first compute the interior and exterior spacetime of uniformly rotating NSs in order to derive the equations of motion for the test-particle.  Following \cite{2015PhRvD..92b3007C}, we consider the stationary axisymmetric spacetime metric in quasi-isotropic coordinates and in geometric units $c=G=1$ \citep{1994ApJ...424..823C}:
\begin{eqnarray}
\label{eq1}
ds^2 &=& -e^{\gamma + \rho} dt^2 + e^{\gamma-\rho} r^2 \sin^2 \theta (d \phi - \omega dt)^2 + \nonumber \\
         &+& e^{2 \lambda} (dr^2 + r^2 d \theta^2),
\end{eqnarray}
where $\gamma$, $\rho$, $\omega$ and $\lambda$ depend only on variables $r$ and $\theta$.

It is useful to introduce the variable $e^{\psi} =  r \sin(\theta) B e^{-\nu}$, being again $B = B(r,\theta)$. The energy-momentum tensor of the NS interior is given by
\begin{equation}
T^{\alpha \beta} = (\varepsilon + P) u^\alpha u^\beta + P g^{\alpha \beta},
\end{equation}
where $\varepsilon$ and $P$ denote the energy density and pressure of the fluid, and $u^\alpha$ is the fluid 4-velocity. Thus, with the metric given by equation~(\ref{eq1}) and the above energy-momentum tensor, one can write the field equations as (setting $\zeta=\lambda + \nu$):
\begin{widetext}
\begin{subequations}
\begin{eqnarray}
\label{eqA1a}
&&\nabla\cdot\left( B \nabla \nu  \right) = \frac{1}{2} r^2 \sin^2 \theta B^3 e^{-4\nu} \nabla \omega \cdot \nabla \omega
+ 4 \pi B e^{2\zeta - 2\nu} \left[ \frac{(\varepsilon+P)(1+v^2)}{1-v^2} + 2P \right],\\
\label{eqA1b}
&&\nabla \cdot \left(  r^2 \sin^2 \theta B^3 e^{-4\nu} \nabla \omega \right) = -16 \pi r \sin \theta B^2 e^{2\zeta - 4\nu} \frac{(\varepsilon +P)v}{1-v^2},
\\
\label{eqA1c}
&&\nabla \cdot \left( r \sin(\theta) \nabla B \right) = 16 \pi r \sin \theta B e^{2\zeta - 2\nu} P,
\end{eqnarray}
\begin{eqnarray}
\label{eqA1d}
{\zeta}_{,\mu} =& - & {\left\lbrace \left( 1-{\mu}^2 \right) {\left( 1+r \frac{B_{,r}}{B} \right)}^2 + {\left[ \mu - \left( 1-{\mu}^2 \right) \frac{B_{,r}}{B} \right]}^2 \right\rbrace}^{-1}  \Biggl[\frac{1}{2} B^{-1} \left\lbrace r^2 B_{,rr} - {\left[ \left( 1 -{\mu}^2 \right) B_{,\mu} 
                       \right]}_{,\mu} - 2\mu B_{,\mu} \right\rbrace  \nonumber \\%
                       & \times & \left\lbrace - \mu + \left( 1-{\mu}^2 \right) \frac{B_{,\mu}}{B} \right\rbrace +  r\frac{B_{,r}}{B} \left[ \frac{1}{2} \mu + \mu r\frac{B_{,r}}{B} + \frac{1}{2} \left( 1-{\mu}^2 \right) \frac{B_{,\mu}}{B} \right] + \frac{3}{2} \frac{B_{,\mu}}{B} \left[ -{\mu}^2 + \mu \left( 1-{\mu}^2 \right) \frac{B_{,\mu}}{B} \right]\nonumber\\%
                       & - & \left( 1-{\mu}^2 \right) r\frac{B_{,\mu r}}{B} \left( 1+r\frac{B_{,r}}{B} \right) - \mu r^2 {\left( {\nu}_{,r} \right)}^2 - 2\left( 1-{\mu}^2 \right) r{\nu}_{,\mu}{\nu}_{,r} + \mu \left( 1-{\mu}^2 \right){\left( {\nu}_{,\mu} \right)}^2 - 2\left( 1-{\mu}^2 \right) r^2 B^{-1} B_{,r} {\nu}_{,\mu} {\nu}_{,r}\nonumber \\%
                       & + & \left( 1-{\mu}^2 \right) B^{-1} B_{,\mu} \left[ r^2 {\left( {\nu}_{,r} \right)}^2 - \left( 1-{\mu}^2 \right) {\left( {\nu}_{, \mu} \right)}^2 \right] + \left( 1-{\mu}^2 \right) B^2 e^{- 4\nu} \Bigl\lbrace \frac{1}{4} \mu r^4 {\left( {\omega}_{,r} \right)}^2 + \frac{1}{2} \left( 1-{\mu}^2 \right) r^3 {\omega}_{,\mu} {\omega}_{,r}\nonumber \\%
                       & - & \frac{1}{4} \mu \left( 1-{\mu}^2 \right) r^2 {\left( {\omega}_{, \mu} \right)}^2 + \frac{1}{2} \left( 1-{\mu}^2 \right) r^4 B^{-1} B_{,r} {\omega}_{,\mu} {\omega}_{,r} - \frac{1}{4} \left( 1-{\mu}^2 \right) r^2B^{-1}B_{,\mu} \left[ r^2 {\left( {\omega}_{,r} \right)}^2 - \left( \-{\mu}^2 \right) {\left( {\omega}_{,\mu} \right)}^2 \right] \Bigr\rbrace \Biggr],
\end{eqnarray}
\end{subequations}
\end{widetext}
where, in the equation for ${\zeta}_{,\mu}$, we introduced $\mu\equiv \cos(\theta)$. 

The NS interior is made of a core and a crust. The core of the star has densities higher than the nuclear value, $\rho_{\rm nuc}\approx 3\times 10^{14}$~g~cm$^{-3}$, and it is composed by a degenerate gas of baryons (e.g.~neutrons, protons, hyperons) and leptons (e.g.~electrons and muons). The crust, in its outer region ($\rho \leq \rho_{\rm drip}\approx 4.3\times 10^{11}$~g~cm$^{-3}$), is composed of ions and electrons, and in the so-called inner crust ($\rho_{\rm drip}<\rho<\rho_{\rm nuc}$), there are also free neutrons that drip out from the nuclei. For the crust, we adopt the Baym-Pethick-Sutherland (BPS) EOS \citep{1971ApJ...170..299B}. For the core, we  adopt instead relativistic mean-field (RMF) theory models within the extension of the formulation of Boguta \& Bodmer \cite{1977NuPhA.292..413B} with massive scalar and vector meson mediators ($\sigma$, $\omega$, and $\rho$ mesons). In this work we present results for NSs constructed using the NL3 \cite{1997PhRvC..55..540L}, TM1 \cite{1994NuPhA.579..557S} and GM1 \cite{1991PhRvL..67.2414G,2000NuPhA.674..553P} EOS. 

\begin{figure}
\centering
\includegraphics[width=\hsize,clip]{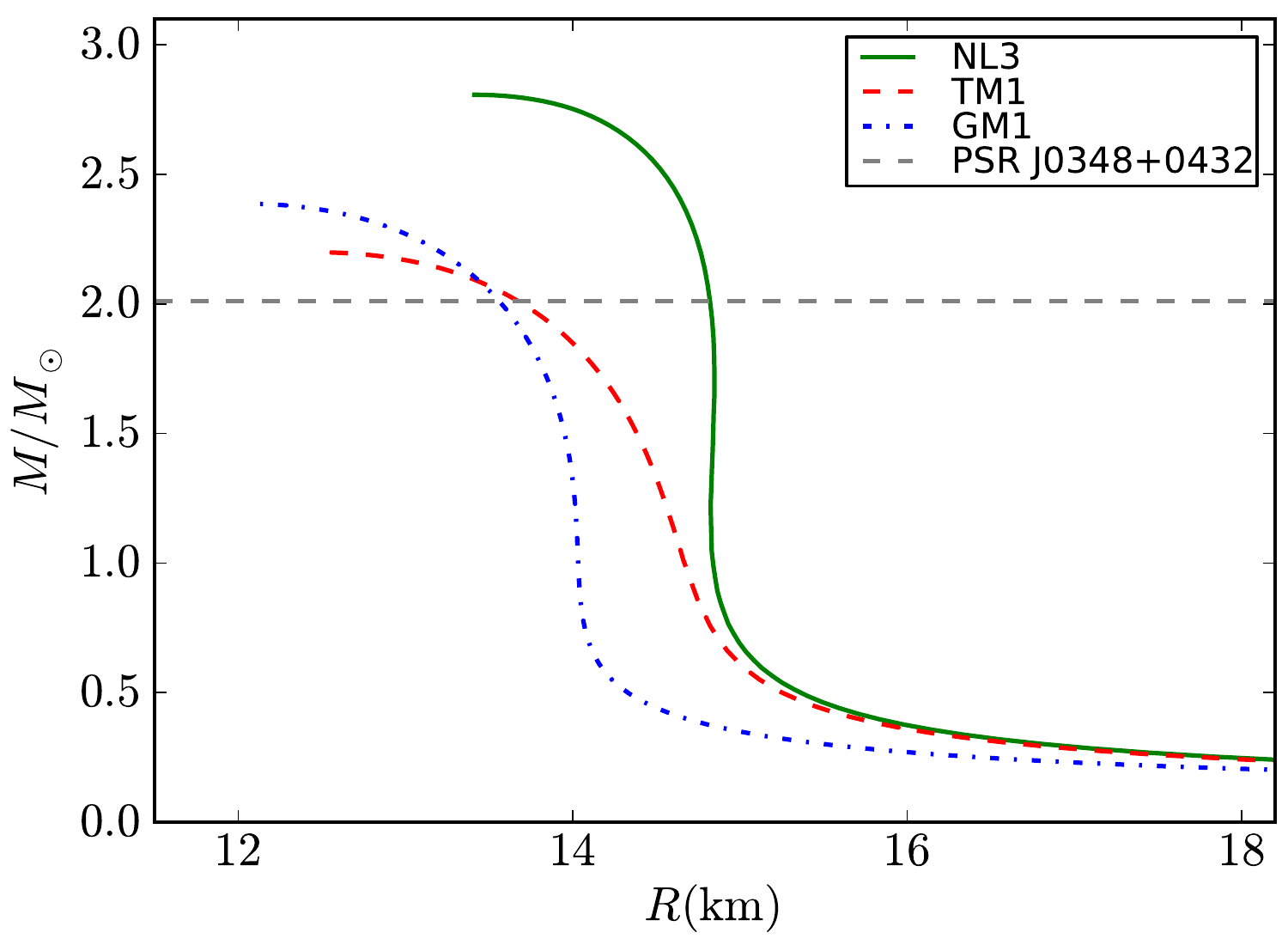}
\caption{Mass-radius relation for non-rotating NSs for the three EOS NL3 (green solid curve), TM1 (red dashed curve) and GM1 (blue dotted-dashed curve) used in this work. The gray dashed horizontal line shows the mass of the heaviest NS observed, PSR J0348+0432, $M = 2.01 \pm 0.04 M_\odot$ \cite{2013Sci...340..448A}.}
\label{fig:MR}
\end{figure}

Our preference for EOS based on RMF models is because they satisfy important properties such as Lorentz covariance, they are self-consistent relaivistic models and therefore they do not violate causality, and they are successful in providing an intrinsic inclusion of spin as well as a simple mechanism of saturation of nuclear matter. We refer to Refs.~\cite{PhysRevC.90.055203,PhysRevC.93.025806} for recent extensive studies of RMF models both from the nuclear experiments point of view and from the astrophysical one. The above three representative models that we use in this work satisfy the astrophysical constraint of producing non-rotating, stable NSs up to masses larger than the most massive NS observed, PSR J0348+0432, with $M = 2.01 \pm 0.04 M_\odot$ \cite{2013Sci...340..448A}. The mass-radius relation for non-rotating models obtained with these three EOS is shown in Fig.~\ref{fig:MR}.

With the knowledge of the EOS we can compute equilibrium configurations integrating the above Einstein equations for suitable initial conditions, e.g.~central density and angular momentum (or angular velocity) of the star. Then the properties of the NS such as the total gravitational mass, the total baryon mass, polar and equatorial radii, moment of inertia, quadrupole moment, etc, can be obtained as a function of the central density and angular momentum. 

The equilibrium configurations are limited by the Keplerian, mass-shedding, or maximally rotating sequence, and by the secular axisymmetric instability. At the Keplerian sequence the dimensionless angular momentum $a/M\equiv c J/(G M^2)$, where $a=J/M$ is the angular momentum per-unit-mass, reaches a maximum value of $a_{\rm max}/M\approx 0.7$, independently on the EOS \cite{2015PhRvD..92b3007C}. This value is lower than the maximum dimensionless angular momentum parameter of a rotating BH given by the extreme Kerr solution, i.e.~$a_{\rm max, BH}/M_{\rm BH}=1$.

The secular axisymmetric instability sequence separates stable from unstable stars against axisymmetric perturbations. The turning-point method \cite{1988ApJ...325..722F} gives a sufficient condition for the onset of this instability. Such a sequence, for the present EOS, is well fitted by
\begin{equation}\label{eq:Mcrit}
M_{\rm NS}^{\rm crit}=M_{\rm crit}^{J=0}(1 + k j_{\rm NS}^p),
\end{equation}
with a maximum error of 0.45\% \cite{2015PhRvD..92b3007C}. The parameters $k$ and $p$ and $M_{\rm crit}^{J=0}$ depend of the nuclear EOS (see table \ref{tb:StaticRotatingNS}). The latter, the critical NS mass in the non-rotating case, is as expeceted below the upper bound to the critical mass by Rhoades \& Ruffini, $3.2~M_\odot$ \cite{1974PhRvL..32..324R}.
\begin{table}
\centering
\begin{tabular}{cccc}
\hline \hline
EOS  &  $M_{\rm crit}^{J=0}$~$(M_{\odot})$ &$p$&$k$ \\ 
\hline
NL3 & $2.81$ & $1.68$ & $0.006$\\
GM1 & $2.39$ & $1.69$ & $0.011$ \\
TM1 & $2.20$ & $1.61$ & $0.017$\\  
\hline
\end{tabular}
\caption{Parameters needed to compute the secular axisymmetric instability sequence as given by Eq.~(\ref{eq:Mcrit}).}\label{tb:StaticRotatingNS}
\end{table}

\section{Last stable circular orbit}\label{sec:3}

We are interested in circular orbits of particles on the equatorial plane of the NS, that is to fix $\theta = \frac{\pi}{2}$ (see \cite{1970ApJ...161..103B}). It is well-known that a practical way to analyze the problem of the circular orbits is through the effective potential $V(r, \tilde{E}, \tilde{L})$ (see, e.g., Ruffini \& Wheeler 1969 in \S 104 in \cite{1975ctf..book.....L}; see also Refs.~\cite{1972bhgw.book.....R, 1973blho.conf..215B}), whose turning points give us the radii of the circular orbits. For the metric given by equation~(\ref{eq1}), one can express the effective potential $V(r, \tilde{E}, \tilde{L})$ as follows \cite{1994ApJ...424..823C}:
\begin{eqnarray}
\label{eq2}
V(r, \tilde{E}, \tilde{L}) &=& e^{2 \lambda +\gamma} \left( \frac{dr}{d \tau} \right)^2 \nonumber\\
&=& e^{-\rho} \left( \tilde{E} -\omega \tilde{L} \right)^2 - e^{\gamma} - \frac{e^{\rho}}{r^2} {\tilde{L}}^2,
\end{eqnarray}
where $\tau$ is the proper-time of the free particle. In order to obtain a circular orbit, one should impose the conditions
\begin{equation}
\label{eq3}
V = V_{,r} = 0,
\end{equation}
and from equations~(\ref{eq2}) and (\ref{eq3}), one obtains:
\begin{eqnarray}
\label{eq4}
\tilde{E} &=& \frac{\tilde{v} e^{\frac{\gamma+\rho}{2}}}{(1 - {\tilde{v}}^2)^{\frac{1}{2}}} + \omega \tilde{L},
\\
\label{eq5}
\tilde{L} &=& \frac{\tilde{v} r e^{\frac{\gamma-\rho}{2}}}{(1 - {\tilde{v}}^2)^{\frac{1}{2}}},
\end{eqnarray}
with $\tilde{v}$ the velocity as measured by the zero angular momentum observer (ZAMO):
\begin{eqnarray}
\label{eq6}
\tilde{v}  &=& \frac{1}{2 + r(\gamma_{,r}-\rho_{,r})}  \left\lbrace  e^{-\rho} r^2 \omega_{,r} \pm \left[ e^{-2\rho} r^4 {\omega_{,r}}^2 + \right. \right. \nonumber \\
             &+& \left. \left. 2 r(\gamma_{,r}+\rho_{,r}) + r^2({\gamma_{,r}}^2-{\rho_{,r}}^2) \right]^{\frac{1}{2}} \right\rbrace,
\end{eqnarray}
where the upper (plus) sign is for co-rotating particles and the lower (minus) sign is for counter-rotating particles.

Stable orbits are those for which the above equations are satisfied and, in addition, $V_{,rr} \geq 0$, where the equality corresponds to the LSO. We shall denote the radius of the LSO to as $r_{\rm lso}$. Depending upon the mass and angular momentum of the NS, we have situations in which $r_{\rm lso}>r_{\rm eq}$, being $r_{\rm eq}$ the coordinate equatorial radius of the star, and situations in which stable circular orbits exist down to the stellar surface, namely $r_{\rm lso}=r_{\rm eq}$.

\subsection{Location of the last stable circular orbit}\label{sec:3A}

We now check the conditions under which the LSO actually resides outside the NS. It is then clear that, the condition of the LSO to lie outside the NS, i.e. the condition $r_{\rm lso} \geq r_{\rm eq}$, establishes a minimum mass (for a given value of the angular momentum), or conversely, a maximum angular momentum (for a given mass), over which this condition is satisfied. In the case $J=0$, namely for non-rotating stars, it is known that the LSO is located at $r^{J=0}_{\rm lso}=6 G M/c^2$, and therefore the minimum mass to have this orbit outside the star is obtained for the configuration with radius $R=r^{J=0}_{\rm lso}$. For the NL3, TM1 and GM1 EOS, in the case of co-rotating particles this minimum mass is $\left[ 1.68, 1.61, 1.57 \right]~M_\odot$, respectively. On the other hand, for counter-rotating particles, this minimum mass is given for the maximally rotating (Keplerian) configuration and for the NL3, TM1 and GM1 EOS is $\left[ 1.42, 1.41, 1.34 \right]~M_\odot$.

\begin{figure}
\centering
\includegraphics[width=\hsize,clip]{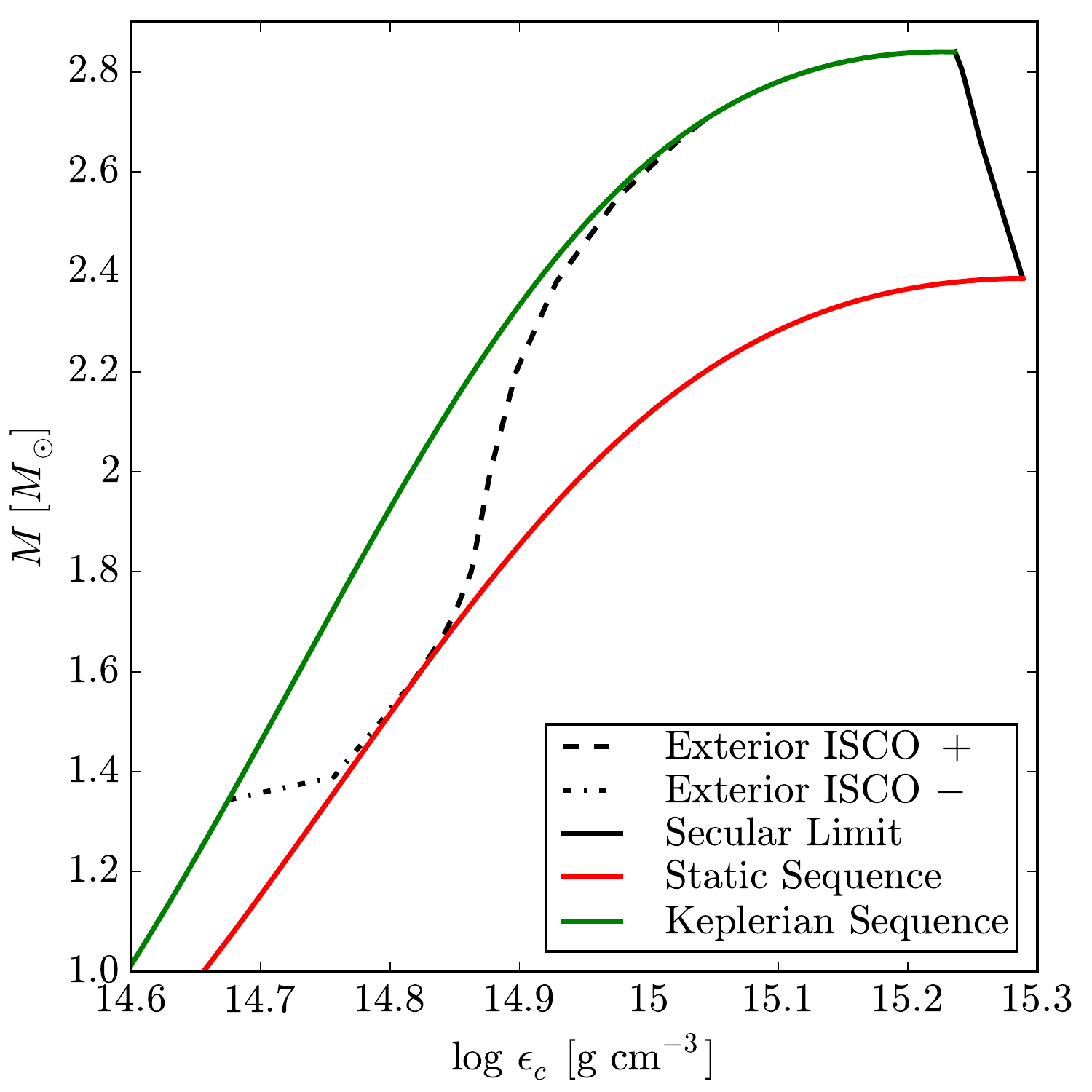}
\caption{Mass versus central density of uniformly rotating NSs with the GM1 EOS. The region of stability is bounded by the non-rotating sequence (solid red curve), the maximally rotating models (solid green curve), namely the mass-shedding limit or Keplerian sequence, and the secular axisymmetric stability limit (solid black curve). The dashed curve corresponds to configurations for which the LSO of co-rotating particles equals the NS equatorial radius: configurations on the right side of it possess an LSO exterior to their surface while configurations on the left side of the curve, have stable circular orbits down to the NS surface. Analogously, the dashed-dotted curve corresponds to configurations for which the LSO of counter-rotating particles equals the NS equatorial radius: configurations above it possess an LSO exterior to their surface while, configurations below it, have stable circular orbits down to the NS surface.}\label{stabplusex}
\end{figure}

Fig.~\ref{stabplusex} shows the results in the rotating case for the GM1 EOS and for co-rotating and counter-rotating orbits. The stable NS models reside in the interior region bounded by the static (solid red curve), Keplerian (solid green curve), and secular instability (solid black curve) sequences. The configurations along the dashed curve have the radius of the LSO equal to the NS equatorial radius, i.e. $r_{\rm lso}=R_{\rm eq}$. Only the configurations on the right side of this curve have $r_{\rm lso} > R_{\rm eq}$. The configurations on the left side of the curve have stable circular orbits down to the NS surface. The dashed-dotted curve is the analogous limit for orbits of counter-rotating particles, thus the configurations under this curve have stable circular orbits down to the NS surface, while the configurations above it have $r_{\rm lso} > R_{\rm eq}$.

For the co-rotating case we can obtain a fitting function of the minimum NS mass, $M_{\rm min}$, for which given a value of the angular momentum one has $r_{\rm LSO}\geq R_{\rm eq}$. For the selected EOS such a function is:
\begin{equation}\label{eq:fitjmax}
\frac{M_{\rm min}}{M_\odot} = \frac{M_{\rm min}^{j=0}}{M_\odot} + c_1 j^{c_2},
\end{equation}
where $M_{\rm min}^{j=0}$, $c_{1}$ and $c_{2}$ are dimensionless constants that depend on the EOS. We report the values of these fitting parameters in Table~\ref{tab:FITNS} together the maximum relative error and the values of NS mass for which this maximum error is obtained. Clearly, the above fitting formula is valid up to the configuration that intersects the Keplerian sequence, namely where the dashed black curve intersects the solid green curve in Fig.~\ref{stabplusex}. The value of the dimensionless angular momentum of that configuration, which we denote here to as $j_{\rm max}$, is reported in Table~\ref{tab:FITNS}. It can be easily checked that introducing the value of $j_{\rm max}$ given in Table~\ref{tab:FITNS} into the Eq.~(\ref{eq:fitjmax}), one obtains the correct value of the mass of this precise configuration on the Keplerian sequence.

%

\begin{table}
\centering
\begin{tabular}{cccccc}
\hline
EOS & $M_{\rm min}^{j=0}\,[M_\odot]$ & $c_{1}$ & $c_{2}$ & ${\rm Max \, rel \, err} (\%)$ & $j_{\rm max}$\\
\hline
NL3 & 1.68 & 0.225 & 0.94 & 1.71 & 6.31\\
TM1 & 1.61 & 0.238 & 0.94 & 1.68 & 4.47\\
GM1 & 1.57 & 0.242 & 0.94 & 1.66 & 4.98\\
\hline
\end{tabular}
\caption{Parameters of the fitting formulas given by Eq.~(\ref{eq:fitjmax}) for the three EOS used, together with maximum relative errors.}\label{tab:FITNS}
\end{table}

It is important to stress that Eq.~(\ref{eq:fitjmax}) is not EOS-independent and it is here presented with the only purpose of providing the reader a complete set of analytic formulas that simplify the analysis of several astrophysical scenarios. The information provided by Eq.~(\ref{eq:fitjmax}) is therefore complementary to the one recalled in Sec.~\ref{sec:1} on the NS binding energy and accretion luminosity, and the one on the LSO energy and angular momentum that is obtained in the next Sec.~\ref{sec:5}.

\subsection{Orbital binding energy and angular momentum}\label{sec:3B}

We now focus on the properties of the LSO, therefore we deal with NS configurations with $r_{\rm LSO}\geq R_{\rm eq}$. We here present the numerical results obtained through integrations performed with RNS public code (http://www.gravity.phys.uwm.edu/rns/) for NSs  considering mass-constant sequences within the region of stability bounded by the spherical symmetric case (non-rotating), by the Keplerian sequence (mass-shedding) and by the secular axisymmetric instability limit. We shall refer to as supramassive NSs those with a mass larger than the critical mass of non-rotating NSs, i.e. configurations without a stable non-rotating counterpart.

\begin{figure}
\centering
\includegraphics[width=\hsize,clip]{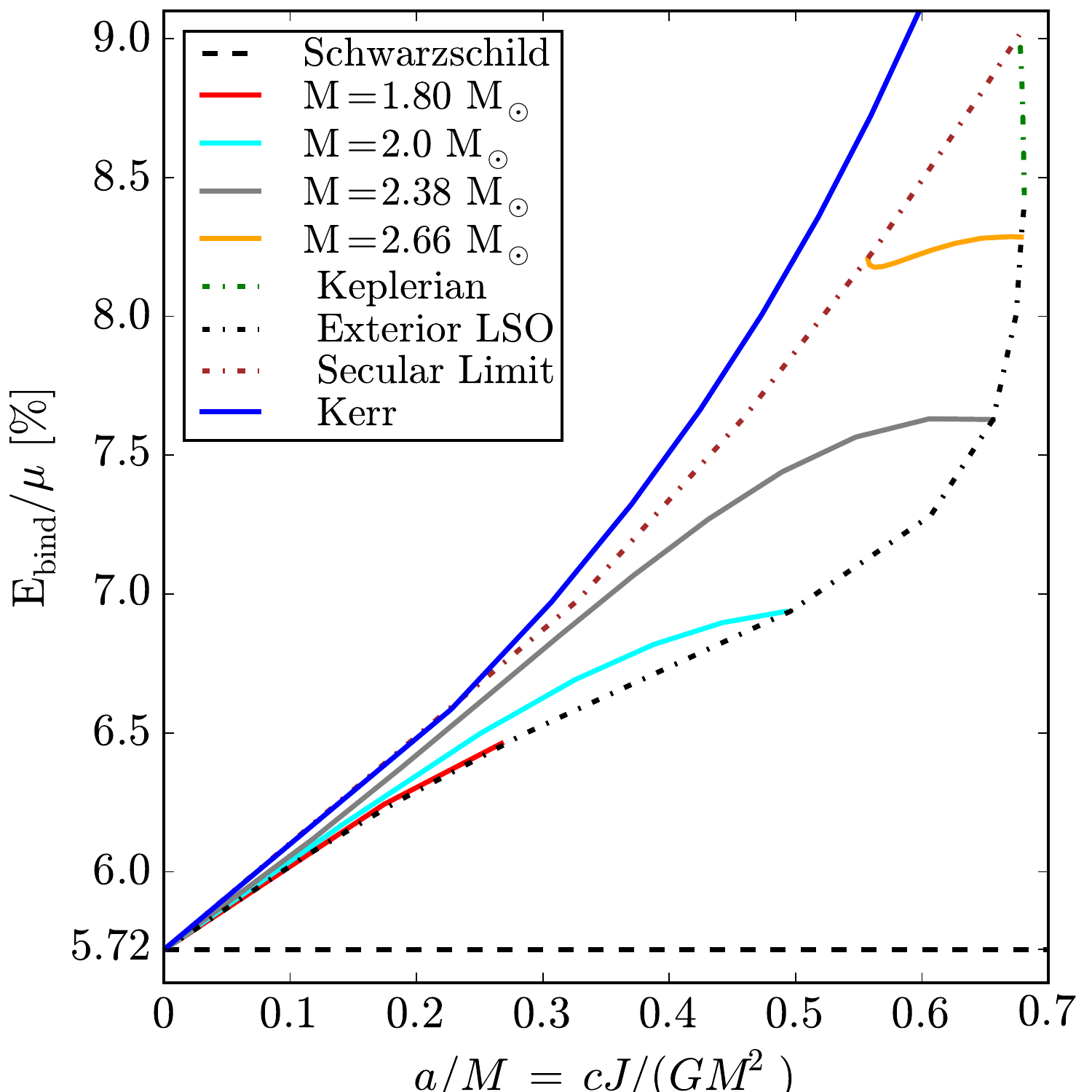}
\caption{Binding energy ($E_{\rm bind}/\mu \equiv 1-\tilde{E}$) of co-rotating test-particles in the LSO for constant mass sequences of NS configurations versus the dimensionless angular momentum $a/M=c J/(G M^2)$. We compare and contrast our results with the values given by the Schwarzschild and Kerr solutions. In this example the NSs obey the GM1 EOS.}\label{GM1_Ebind_p}
\end{figure}

\begin{figure}
\centering
\includegraphics[width=\hsize,clip]{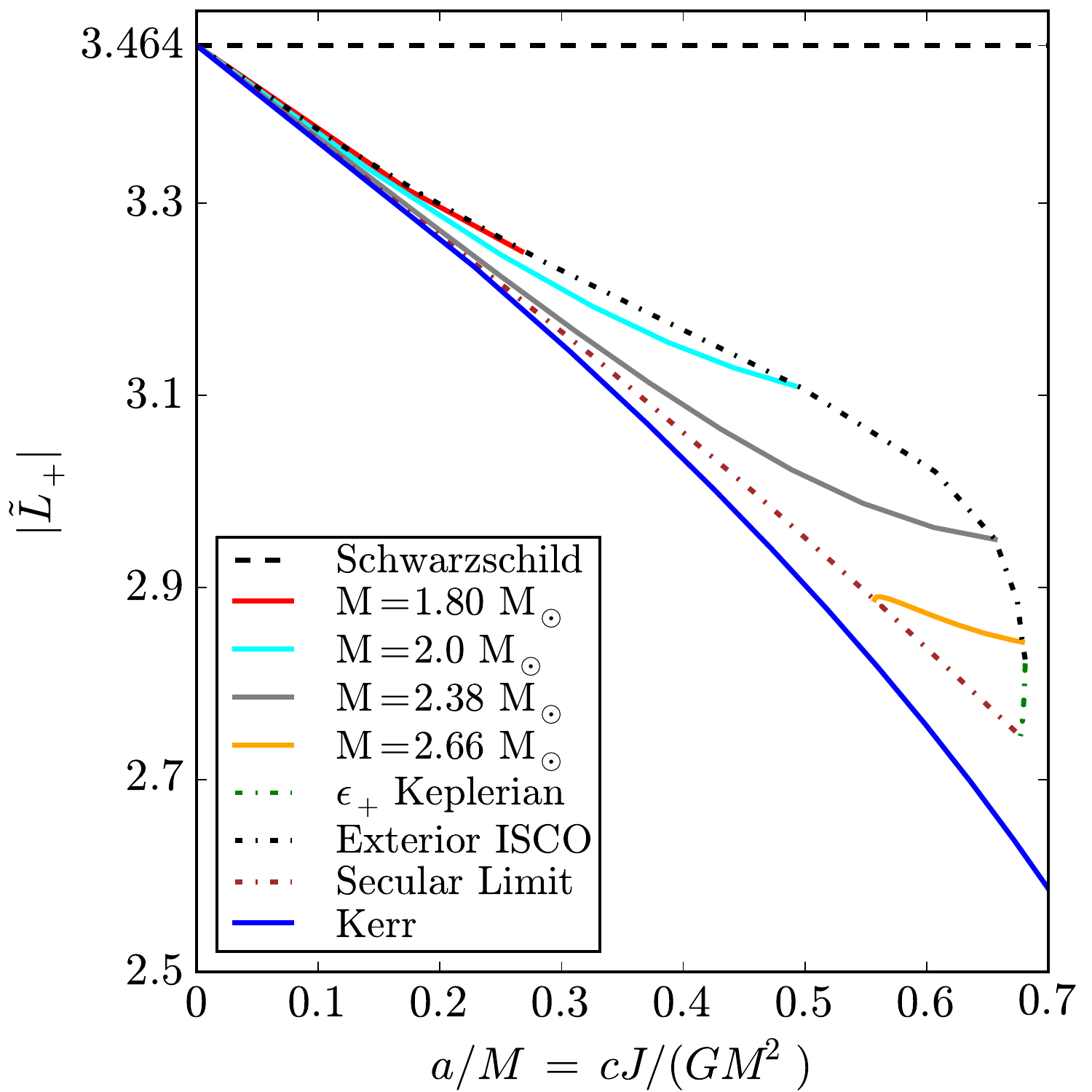}
\caption{Dimensionless angular momentum ($|L|/(\mu M)$) of co-rotating test-particles in the LSO for constant mass sequences of NS configurations versus the dimensionless angular momentum $a/M=c J/(G M^2)$. We compare and contrast our results with the values given by the Schwarzschild and Kerr solutions. In this example the NSs obey the GM1 EOS.}\label{GM1_L_p}
\end{figure}

\begin{figure}
\centering
\includegraphics[width=\hsize,clip]{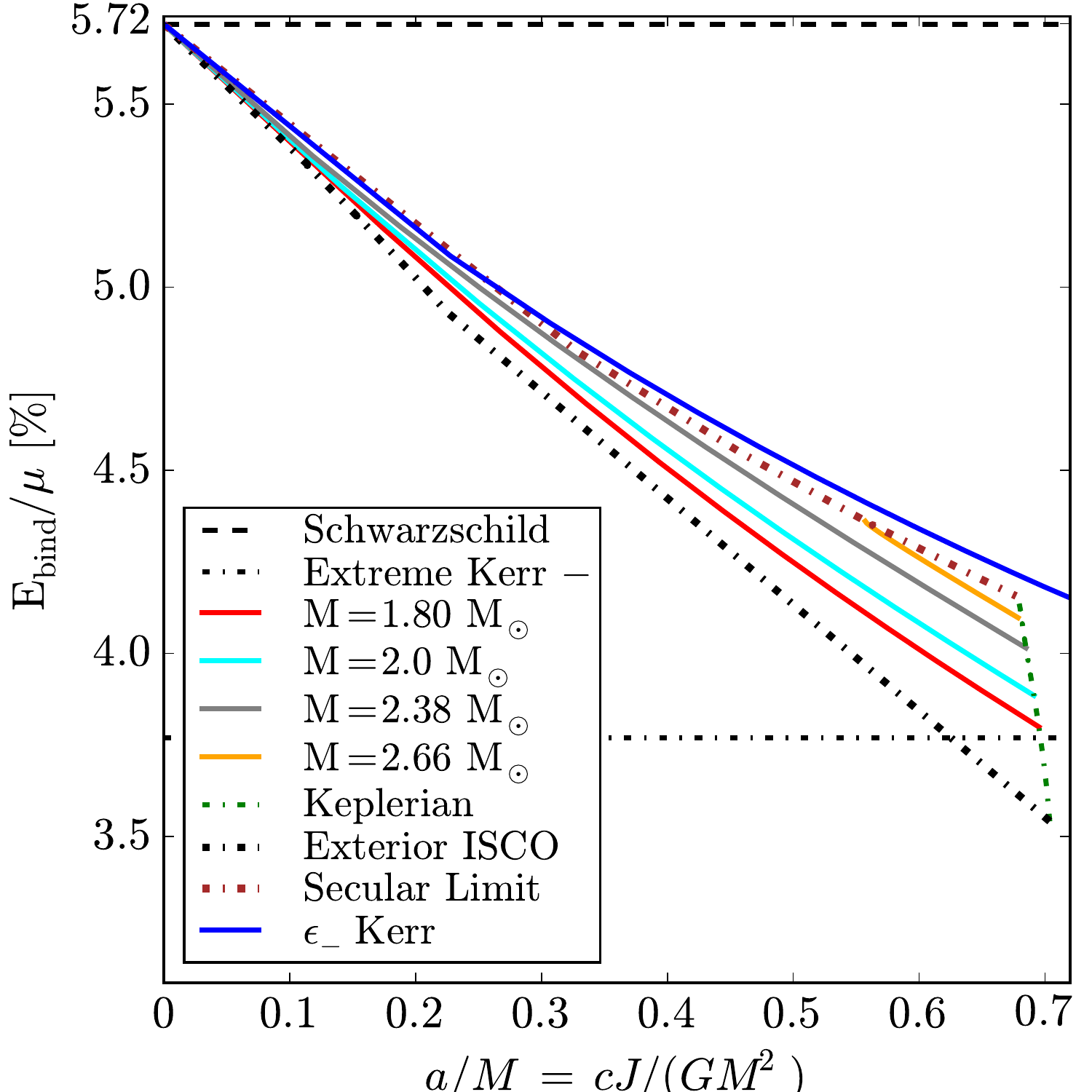}
\caption{Same as Fig.~\ref{GM1_Ebind_p} but for counter-rotating orbits.}\label{GM1_Ebind_m}
\end{figure}

\begin{figure}
\centering
\includegraphics[width=\hsize,clip]{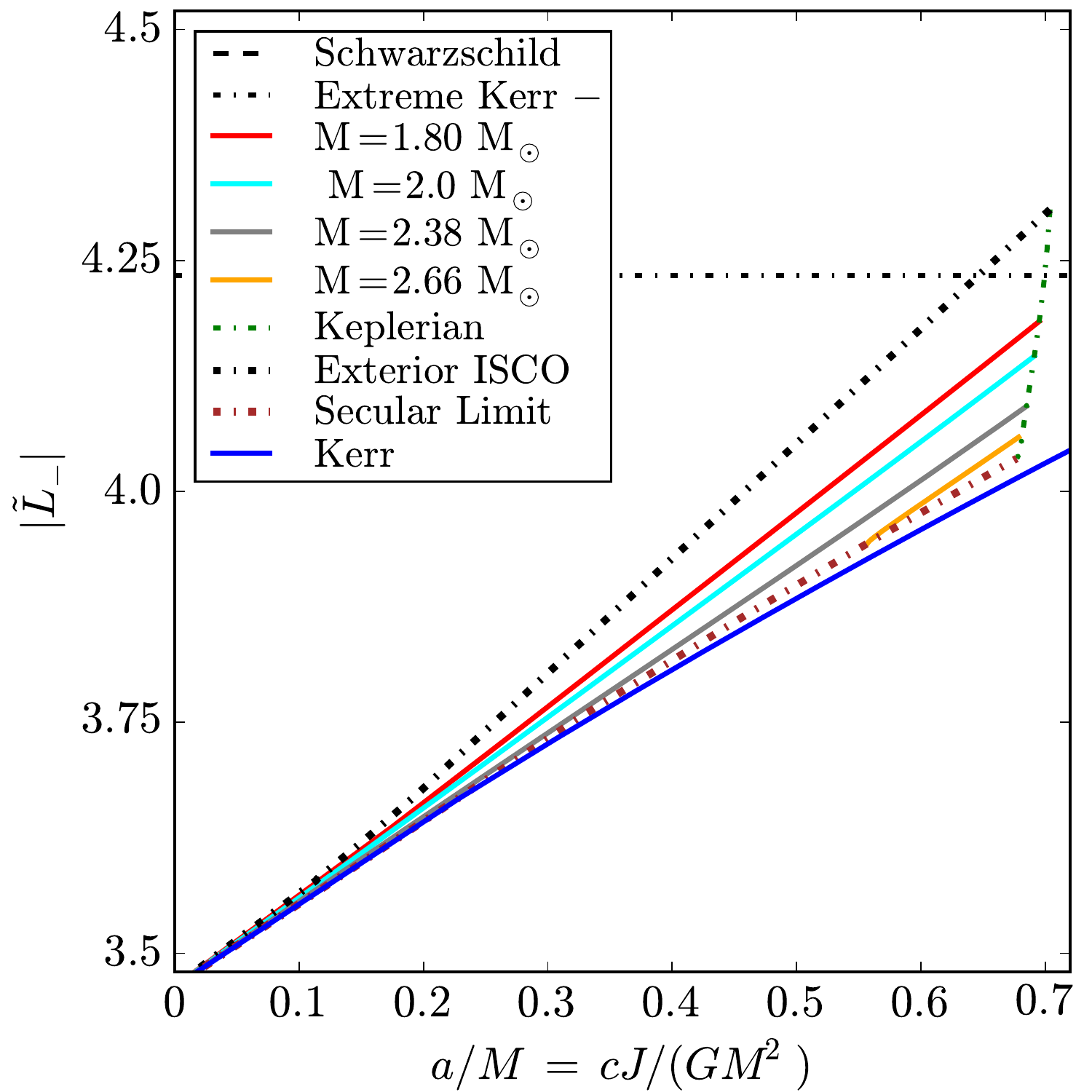}
\caption{Same as Fig.~\ref{GM1_L_m} but for counter-rotating orbits.}\label{GM1_L_m}
\end{figure}

We show in Figs.~\ref{GM1_Ebind_p}--\ref{GM1_L_m} the results of our computations for co-rotating and counter-rotating orbits around NSs obeying the GM1 EOS. The results for the other EOS are analogous. Fig.~\ref{GM1_Ebind_p} shows the binding energy per-unit-mass, $E_{\rm bind}/\mu = 1-\tilde{E}$, as a function of the dimensionless angular momentum parameter, $a/M = cJ/(G M^2)$, for selected constant mass-sequences in case of co-rotating particles. Fig.~\ref{GM1_Ebind_m} shows the results for counter-rotating particles. Fig.~\ref{GM1_L_p} shows the modulus of the dimensionless angular momentum of particles in the LSO, $|L|/(G \mu M/c)$, as a function of $a/M= cJ/(G M^2)$ for the same constant mass sequences in case of co-rotating particles. Fig.~\ref{GM1_L_m} shows the results for counter-rotating particles. 

It can be seen that the sequences are bounded by the Keplerian (mass-shedding) sequence, i.e. $a/M\approx 0.7$, by the limit $r_{\rm LSO}=R_{\rm eq}$, by the secular axisymmetric instability and by the non-rotating limit at $a/M = 0$ (except the supramassive sequences which have no static counterpart), for which the LSO properties have the well-known results of the Schwarzschild exterior solution. We recall that $j = c J/(G M_\odot^2) = (a/M) (M/M_\odot)^2$. We compare and contrast our results with the corresponding values given by the Kerr metric \cite{1973blho.conf..215B}. Deviations from the behavior given by the Kerr solution are evident at almost any value of the dimensionless angular momentum, except for the region of very slow rotation $a/M\ll 1$. 

As one can note from Figs.~\ref{GM1_Ebind_p}, \ref{GM1_L_p}, \ref{GM1_Ebind_m} and \ref{GM1_L_m}, the binding energy and the angular momentum of particles orbiting rotating NSs seem to be power-law functions of the mass and the dimensionless angular momentum. Indeed, we find that the following relations
\begin{eqnarray}
\label{eq10}
\tilde{E} - \tilde{E}_0 &=& \mp 0.0132 \left(\frac{j}{M/M_\odot}\right)^{0.85},
\\
\label{eq11}
|\tilde{L}| - \tilde{L}_0 &=& \mp 0.37 \left(\frac{j}{M/M_\odot}\right)^{0.85},
\end{eqnarray}
where the upper(lower) sign corresponds to co(counter)-rotating orbits, hold for the three studied EOS. This leads to the conjecture that these relations might be universal. The values $\tilde{E}_0=\sqrt{8/9}$ and $\tilde{L}_0 = 2 \sqrt{3}$ are the well-known values of the Schwarzschild solution, hence our formulas recover the correct values in the non-rotating case. We note that in the slow rotation regime, $a/M \ll 1$, the Kerr solution seems to approach this behavior (see Figs.~\ref{GM1_Ebind_p} and \ref{GM1_L_p}, for the co-rotating case). The above fitting formula for $\tilde{E}$ is accurate with a maximum error of 1\% and the one for $\tilde{L}$ has a maximum error of 0.3\%. It is interesting to note that we obtain that the same fitting formulas apply to both co- and counter-rotating orbits.

\begin{figure*}
\centering
\includegraphics[width=0.49\hsize,clip]{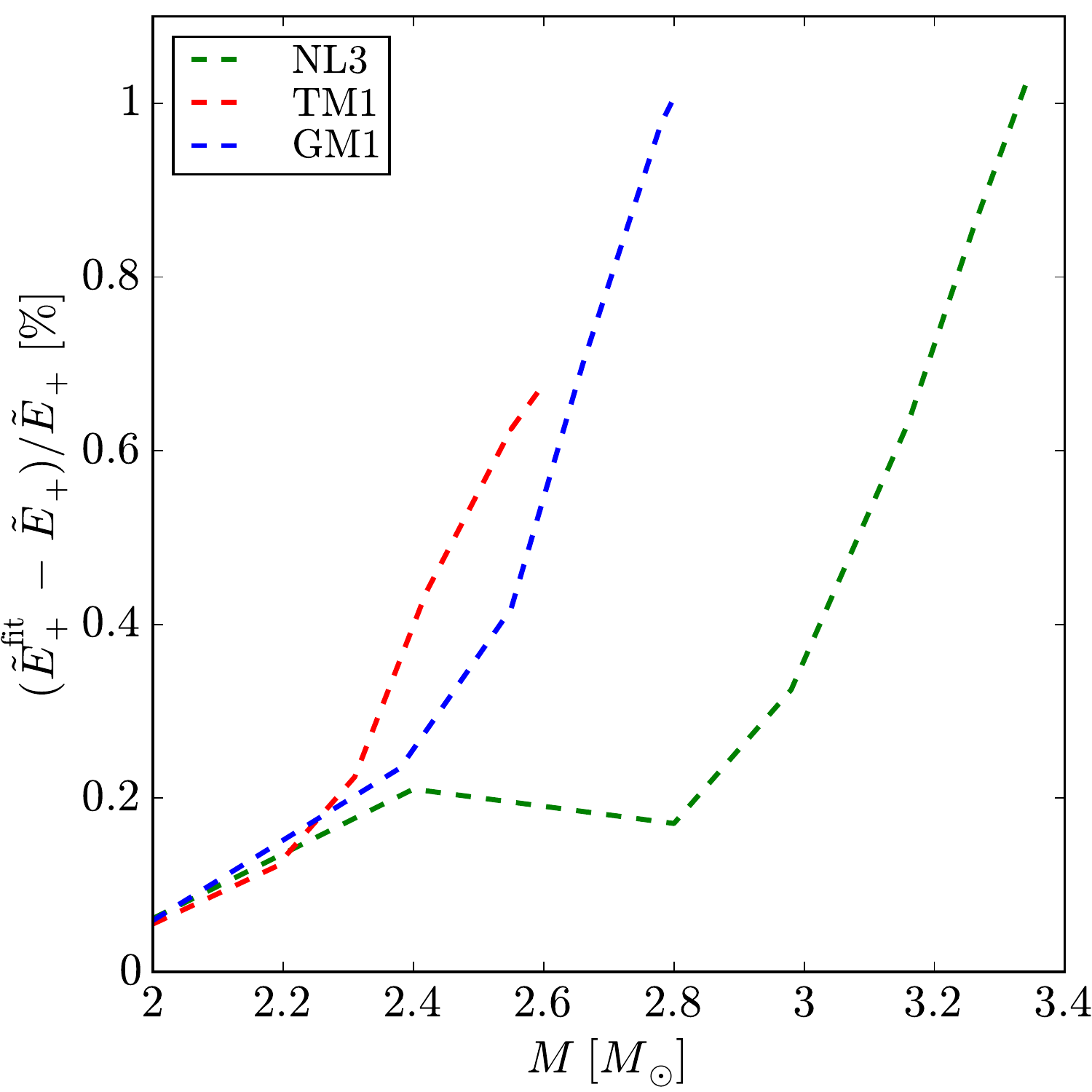}\includegraphics[width=0.49\hsize,clip]{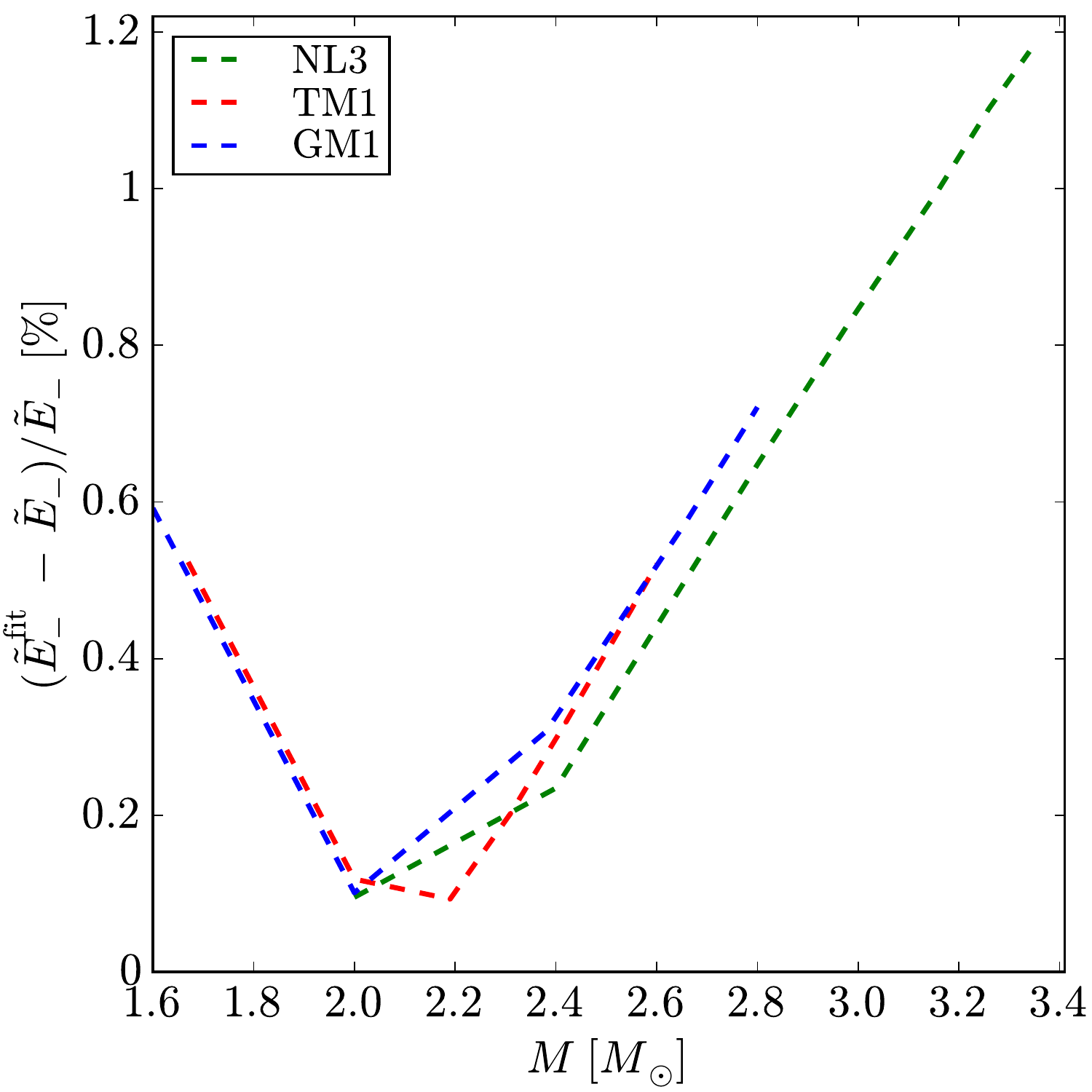}\\
\includegraphics[width=0.49\hsize,clip]{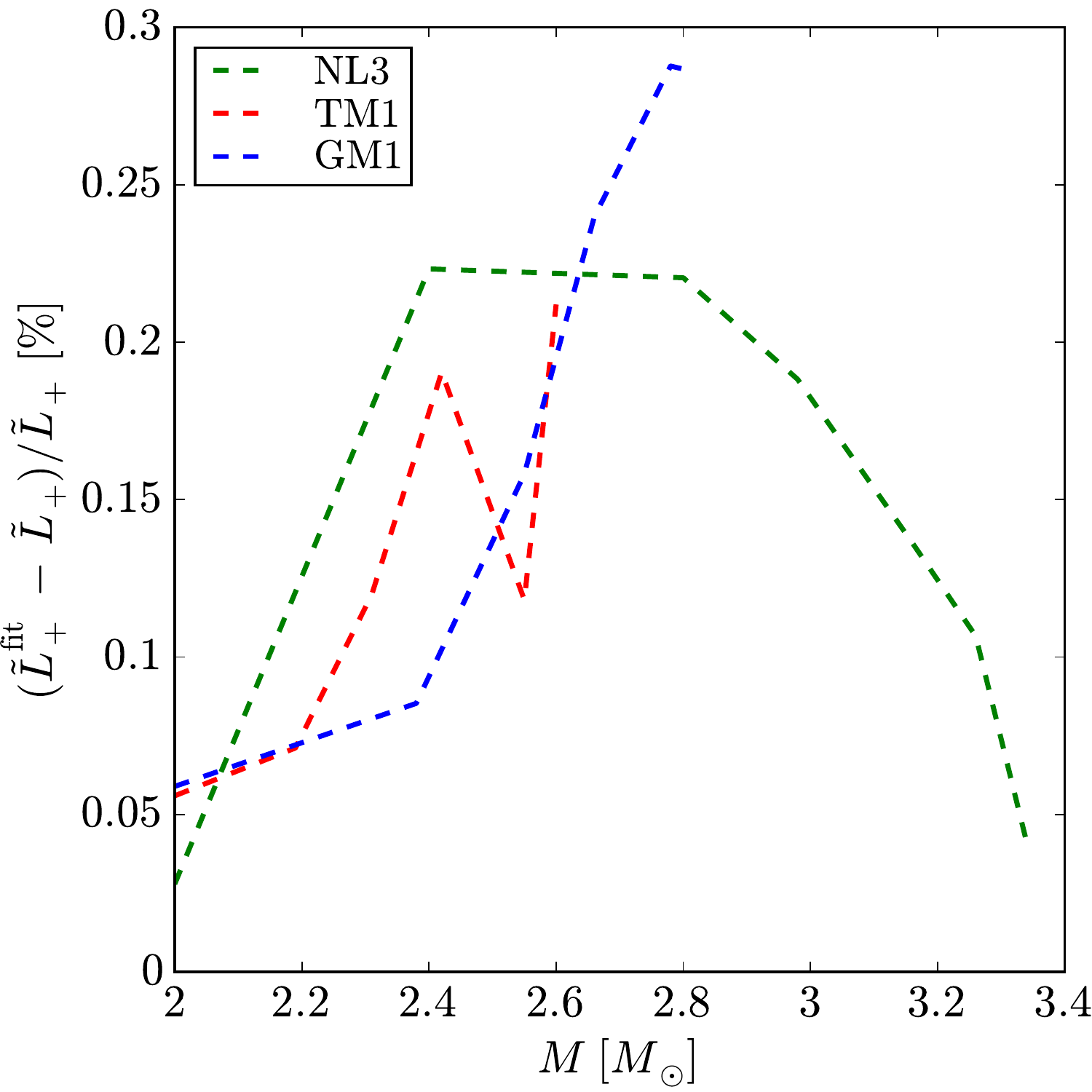}\includegraphics[width=0.49\hsize,clip]{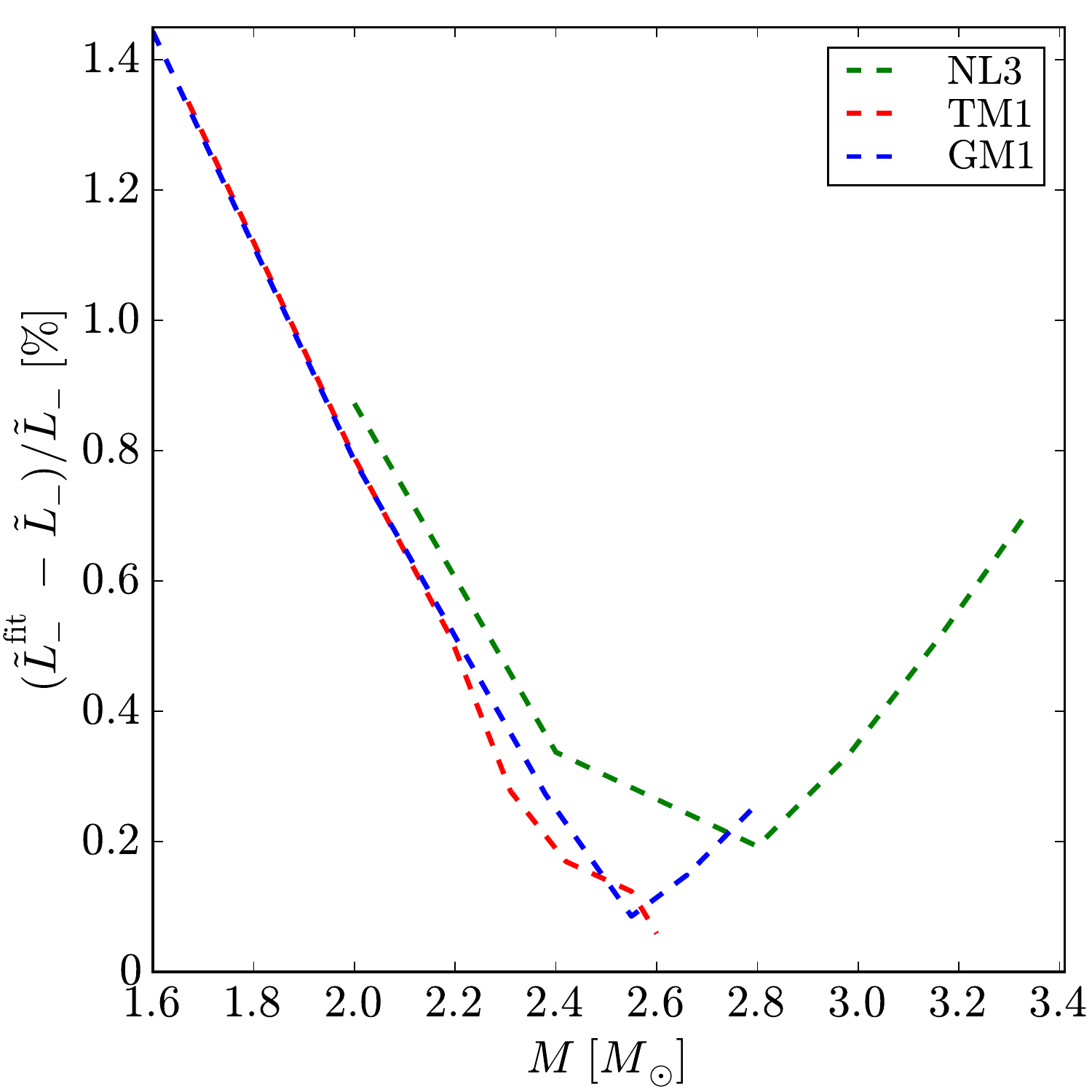}
\caption{Maximum error (in percentage) of Eqs.~(\ref{eq10}) and (\ref{eq11}) with respect to the numerical value of $\tilde{E}$ and $\tilde{L}$ for the sequences of constant gravitational mass in the the range $2$--$3.4~M_\odot$. The results for co-rotating orbits are shown in the left upper and lower panels and for counter-rotating ones in the right upper and lower ones.}
\label{fig:errors}
\end{figure*}

We have shown in Figs.~\ref{GM1_Ebind_p}--\ref{GM1_L_m} the results for the GM1 EOS. For the other EOS similar plots are obtained. Indeed, the formulas (\ref{eq10}) and (\ref{eq11}) perform with similar accuracy in the case of the TM1 and NL3 EOS. In Fig.~\ref{fig:errors} we show the details of the performance of formulas (\ref{eq10}) and (\ref{eq11}) as a function of the NS mass for the three EOS. Specifically, for each sequence of fixed gravitational mass we compute the the maximum error (in percentage) of Eqs.~(\ref{eq10}) and (\ref{eq11}) with respect to the values of $\tilde{E}$ and $\tilde{L}$ obtained from the numerical integration. Fig.~\ref{fig:errors} shows the results for the range of mass $2$--$3.4~M_\odot$ for co- and counter-rotating orbits.

\section{Discussion}\label{sec:4}

We have shown that expressions for $\tilde{E}$ and $\tilde{L}$ remain rather accurate for the three EOS used in this work. One is therefore brought to conjecture on the possible ``universality'' of such equations, namely that such simple relations would remain valid for a broader set of NS EOS. Below, through a set of logically connected statements, we shall conclude that this should be indeed the case.

\begin{enumerate}
\item
There is a firm observational lower limit to the NS critical mass: it must be larger than the mass of the heaviest NS observed, 2.01 $\pm$ 0.04 $M_{\odot}$, of PSR J0348+0432 \cite{2013Sci...340..448A}. 

\item
The above point constraints the nuclear EOS to be stiff. These EOS with sound velocity approaching, but not exceeding, the speed of light (see, e.g., \cite{2014PhRvC..89c5804R}), have a very narrow critical mass \emph{domain of dependence} \cite{1973blho.conf..451R,1974PhRvL..32..324R}.

\item
For such stiff EOS, the condition for the existence of an LSO, namely that the radius of the NS is smaller than the LSO radius, is satisfied only for heavy NSs. In the specific cases studied in this work we have shown that this condition implies $M\gtrsim 1.7~M_\odot$. For details we refer to Sec.~\ref{sec:3A}, specifically to Eq.~(\ref{eq:fitjmax}) with the aid of Table~\ref{tab:FITNS}.

\item
In Ref.~\cite{2000AstL...26..699S}, it was presented a general expansion of the LSO energy $\tilde{E}$ and angular momentum $\tilde{L}$ in terms of $\alpha\equiv a/M=J/M^2$, the NS dimensionless angular momentum parameter, and in terms of the dimensionless quadrupole moment $q\equiv Q/M^3$. Such an expansion shows that the dependence of $\tilde{E}$ and $\tilde{L}$ on the EOS occurs first at linear order in $q$.

\item 
On the other hand, it has been shown that the dimensionless quadrupole moment of NSs can be written as $q=k({\rm EOS},M) \alpha^2$, where the coefficient $k({\rm EOS},M)$ depends on the NS mass and the EOS (see, e.g., Ref.~\cite{1999ApJ...512..282L}). The dependence $q\propto \alpha^2$ is satisfied by both slow and fast rotating NSs. Typically $k>1$ but the larger the NS mass, the more $k$ approaches unity, namely the quadrupole moment of massive NSs approaches the one of the Kerr solution (see Refs.~\cite{2013MNRAS.433.1903U,2015PhRvD..92b3007C} for more details).

\item 
The above points 4 and 5 imply that the dependence $\tilde{E}$ and $\tilde{L}$ on the EOS occurs only at order $\alpha^2$ through the $k$ and, since for NSs $\alpha<0.7$ \cite{2015PhRvD..92b3007C}, such EOS dependence is expected to be weak.

\item
Following Ref.~\cite{2000AstL...26..699S}, we can write up to second order in $\alpha$ (and first order in $q$):
\begin{eqnarray}\label{eq:ELexpansion}
\tilde{E} -\tilde{E}_0 &=& - 0.032 \alpha + \delta E(k) \alpha^2 + {\cal O}(\alpha^3)\\
\tilde{L} -\tilde{L}_0 &=& - 0.943 \alpha + \delta J(k) \alpha^2 + {\cal O}(\alpha^3)
\end{eqnarray}
where $\delta E(k) = 0.008 k -0.022$ and $\delta J(k) = 0.189 k-0.258$. For values of $k$ of the order of unity as the ones expected for the aforementioned massive NSs of points 1--3, both $\delta E(k)$ and $\delta J(k)$ imply a very small deviation of the $\tilde{E}$ and $\tilde{L}$ from a linear dependence $\alpha^1$. To be more precise, the values of $k$ are such that $\delta E(k)$ and $\delta J(k)$ are slightly positive and therefore the contribution at second order has opposite sign with respect to the one at first order and thus when trying a fit with a sole power of $\alpha$ we should expect a power smaller than unity. Indeed, our results summarized by Eqs.~(\ref{eq10}) and (\ref{eq11}) show $\tilde{E}-\tilde{E}_0 \propto \alpha^{0.85}$ and $\tilde{L}-\tilde{L}_0 \propto \alpha^{0.85}$.

\item
At such linear order in $\alpha$, the LSO energy and angular momentum are indeed ``universal'' since they have no EOS dependence up to this order. The dependence on the EOS should be evident only when the contributions of $\delta E(k)$ and $\delta J(k)$ are non-negligible. This happens for instance when $k\sim 10$ which is the case of NSs with $M\lesssim 1.4~M_\odot$. However, such NSs do not satisfy condition imposed by the point 3 unless the EOS is very soft, but in the latter case from the points 1 e 2 such EOS are not of astrophysical relevance.

\end{enumerate}

It is important to stress that, in general, the energy and angular momentum of the LSO depend on the details of the EOS, however, the above points 1--8 imply that our Eqs.~(\ref{eq10}) and (\ref{eq11}) should remain valid for a wide set of EOS, providing they are of astrophysical relevance in the sense of the points 1 and 2. It is only under these conditions that we can consider these formulas as \emph{universal}.

Although the knowledge of the quadrupole moment appear to be relevant for the determination of several NS properties such as the angular velocity and the LSO radius (see, e.g., Ref.~\cite{1998PhRvD..58j4011S}), our results show that its role in the determination of the energy and angular momentum of the LSO can be much less important. The main reason for this is that, besides being the contribution of order $\alpha^2$ naturally small by itself (because $\alpha <0.7$) with respect to the leading order, the contribution of the quadrupole moment via the coefficients $\delta E(k)$ and $\delta J(k)$, is of opposite sign with respect to the one given by the centrifugal potential, almost canceling each other for the relevant NS masses.  This effect confirms for the LSO the results of Ref.~\cite{2010A&A...520A..16B} on the circular orbits around rotating NSs where this feature had been already noticed.

In Sec.~\ref{sec:3B} we have compared and contrasted our results for $\tilde{E}$ and $\tilde{L}$ with the ones of the LSO in the Kerr background characterized with the same mass and angular momentum. We have seen how the properties of the LSO given by the Kerr metric deviate from the ones of NSs except in the slow rotation regime $\alpha = a/M\gg 1$. This is indeed in agreement with the above discussion on the almost linear dependence in $\alpha$ obtained for $\tilde{E}$ and $\tilde{L}$. Indeed, the expansion of these quantities for small $\alpha$ for the Kerr metric coincide at the linear level (see, e.g, Eqs. B3 and B4 in Ref.~\cite{1998PhRvD..58j4011S}) with the above expansion (\ref{eq:ELexpansion}). Thus, $\tilde{E}$ and $\tilde{L}$ for rotating NSs are relatively well represented by the corresponding values of the Kerr metric kept only at linear order in $\alpha$. However, if more terms of the expansion in the Kerr metric (or the full solution) are taken into account, the predictions of the Kerr solution deviate considerably from the realistic NS values as it is shown in Figs.~\ref{GM1_Ebind_p}--\ref{GM1_L_m}.

\section{Concluding remarks}\label{sec:5}

We have computed the binding energy and angular momentum of test-particles orbiting on the equatorial plane of uniformly rotating NSs. The NS equilibrium configurations were constructed for up-to-date nuclear EOS by integrating the Einstein equations in the axially symmetric case. Our study was limited to stable NSs with respect to the the mass-shedding (Keplerian) limit and the secular axisymmetric instability. Our conclusions are as follows.

\begin{enumerate}

\item 
There is a limiting configuration for which the radius of the LSO equals the equatorial radius of the NS (see, e.g., Fig.~\ref{stabplusex}). As an example, we have obtained the fitting function (\ref{eq:fitjmax}) that connects the mass and angular momentum of such a limiting configuration in the case of co-rotating orbits, for the three EOS used in this work. Thus, given a NS mass(angular momentum), Eq.~(\ref{eq:fitjmax}) gives the maximum(minimum) angular momentum(mass) for which $r_{\rm lso} > R_{\rm eq}$. It is important to recall that Eq.~(\ref{eq:fitjmax}) is not a universal, i.e. EOS-independent equation, and thus it must be computed for every EOS. For more details see Sec.~\ref{sec:3A}.

\item 
We obtained simple formulas for the energy and angular momentum of the LSO of co- and counter-rotating test-particles as a function of the NS mass and angular momentum [see, respectively, Eqs.~(\ref{eq10}) and (\ref{eq11})]. We have obtained these formulas for the three EOS studied in this work (NL3, TM1 and GM1) and are valid for any rotation rate within the established stability limits. 

\item
We have argued that such formulas will remain valid for other nuclear EOS which satisfy the astrophysical request of having a critical NS mass larger than $2~M_\odot$ \cite{2013Sci...340..448A}. The EOS-dependent contributions to $\tilde{E}$ and $\tilde{L}$ appear at higher powers of the dimensionless angular momentum parameter $\alpha = a/M$ and are due to the NS mass quadrupole moment. However, such a contribution becomes negligible for massive NSs which are the ones that possess an LSO exterior to their surface. See Sec.~\ref{sec:4} for details on this discussion.

\item
The simplicity and high accuracy of these formulas, which show a maximum error of 1\% and 0.3\% respectively for the energy and angular momentum of co-rotating orbits (see Fig.~\ref{fig:errors}), makes them particularly suitable for astrophysical applications where taking into due account general relativistic effects of rotating NSs are important, e.g. the accretion processes in X-ray binaries (see, e.g., Refs.~\cite{2011MNRAS.413L..47B,2011A&A...536A..87B,2011A&A...536A..92B}) or hypercritical accretion in GRBs (see, e.g., Refs.~\cite{2016ApJ...833..107B,2015ApJ...812..100B}).

\item 
Our results are qualitatively and quantitatively different from the corresponding ones obtained in the Kerr geometry, except in the slow rotation regime $a/M\ll 1$.

\end{enumerate}

\acknowledgements

We would like to thank the referee for the detailed report which have helped us to improve the presentation of our results. It is a pleasure to thank D.~Bini for discussions on the subject of this work. C. C. and S. F. would like to acknowledge GNFM INdAM and ICRANet for partial support. J.A.R. acknowledges partial support of the project N. 3101/GF4 IPC-11, and the target program F.0679 of the Ministry of Education and Science of the Republic of Kazakhstan.

\bibliography{apspaper2}
\bibliographystyle{apsrev4-1}

\end{document}